\documentclass[amsmath,amssymb,prl,hyperlink,twocolumn]{revtex4}

	\usepackage{graphicx}
	\usepackage{soul}
	\usepackage[colorlinks=true,citecolor=blue,linkcolor=magenta]{hyperref}
	\usepackage[usenames]{color}
	\usepackage{amsfonts}
	\usepackage{color}
	\usepackage{booktabs}
	\usepackage{multirow}
	\usepackage{float}
    \usepackage{times}
    \usepackage[english]{babel}

\begin{document}

\title{Directly Measuring a Multiparticle Quantum Wave Function via Quantum Teleportation}
\author{Ming-Cheng Chen$^{1,2}$, Yuan Li$^{1,2}$, Run-Ze Liu$^{1,2}$, Dian Wu$^{1,2}$, Zu-En Su$^{1,2}$, Xi-Lin Wang$^{1,2}$,  Li Li$^{1,2}$, Nai-Le Liu$^{1,2}$, Chao-Yang Lu$^{1,2}$, and Jian-Wei Pan$^{1,2}$, \vspace{0.2cm}}

\affiliation{$^1$ Hefei National Laboratory for Physical Sciences at Microscale and Department of Modern Physics, University of Science and Technology of China, Hefei, Anhui 230026, China}
\affiliation{$^2$ CAS Centre for Excellence and Synergetic Innovation Centre in Quantum Information and Quantum Physics, University of Science and Technology of China, Hefei, Anhui 230026, China.}
\affiliation{}
\date{\today}

\begin{abstract}
We propose a new method to directly measure a general multi-particle quantum wave function, a single matrix element in a multi-particle density matrix, by quantum teleportation.  The density matrix element is embedded in a virtual logical qubit and is nondestructively teleported to a single physical qubit for readout.  We experimentally implement this method to directly measure the wavefunction of a photonic mixed quantum state beyond a single photon using a single observable for the first time. Our method also provides an exponential advantage over the standard quantum state tomography in measurement complexity to fully characterize a sparse multi-particle quantum state.
\end{abstract}
\maketitle

\begin{figure}[t]
	\centering
	\includegraphics[width=0.45\textwidth]{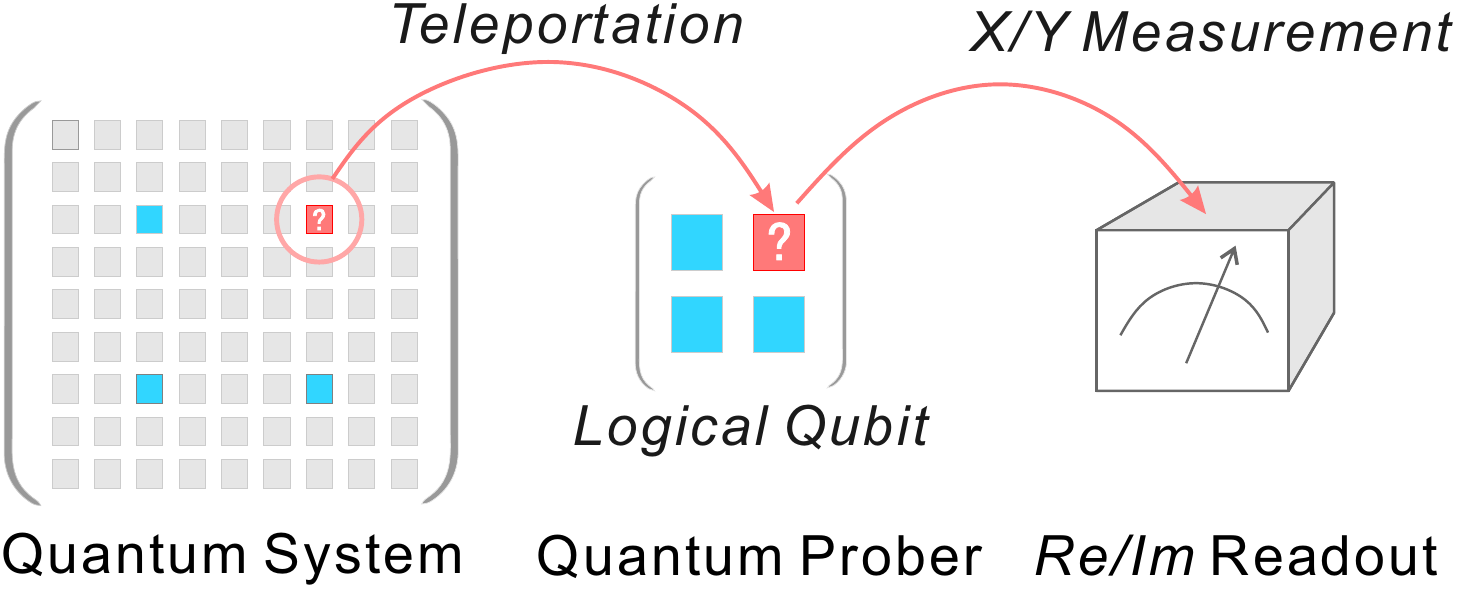}
	\caption{Direct measurement of multi-particle quantum states. An individual density matrix element is embedded in a logical qubit and teleported to a quantum prober. The prober is subsequently measured in $X$ and $Y$ bases to read out the real and imaginary parts of the density matrix element, respectively.}
	\label{fig1}
\end{figure}

Quantum mechanics, based on the underpinning concept of 
complex-valued  quantum wavefunctions, 
is an impressively successful theory of nature \cite{dirac1981}.
Quantum interference in the complex-valued  quantum wavefunction has been exploited to develop novel quantum technology in the past decades, such as quantum computation \cite{nielsen2010} and quantum metrology \cite{metrology}.
However, the fundamental concept of quantum wavefunction is still elusive due to its complex-value nature, which is beyond direct access from practical observation as all physical observables are Hermitian and have real values. 
This renders the wavefunction seems to be an epistemic computing tool, called probability amplitude in the view of Born's rule \cite{born1955}, but not an ontological reality \cite{pusey2012reality}. 
Furthermore, the conventional tomography method used to infer the multi-particle wavefunction needs to collect exponential-scale information of different observables \cite{hradil1997tomo}, 
which has become intractable even for moderate-scale quantum systems \cite{10photons,14ions,10Xmons}. 
Therefore, a direct and efficient method to measure multi-particle wavefunction will not only lead to a better understanding of the fundamental concept of quantum mechanics but also will contribute to characterize the growing experimental quantum systems.

Direct measurement of wavefunction is a concept that uses a single observable to directly obtain the real (or imaginary) part of an individual wavefunction amplitude, in stark contrast to the traditional tomography method. 
It was first related to the weak value in a certain weak measurement process \cite{lundeen2011direct}, which provides a direct correspondence between the abstract wavefunction and practical observables.
This method was further generalized to general mixed quantum states and strong measurements  \cite{lundeen2012procedure,wu2013state,vallone2016strong,zou2015direct} . 
However, these methods demand complicated coherent interaction between the measured system and the pointer of measurement apparatus. For $N$-body quantum system, it needs $N+1$-body Hamiltonian to implement the protocols. This technical obstacle has made the existing experimental realizations limited to only single particle \cite{malik2014direct, shi2015scan, salvail2013full, thekkadath2016direct,directMea2018}. Other technological variants are also developed to experimentally measure two-partite entangled state but using linear combination of several observables and only applied to pure states. \cite{bolduc2016direct,pan2019direct}.

\begin{figure}[tb]
	\centering
	\includegraphics[width=0.46\textwidth]{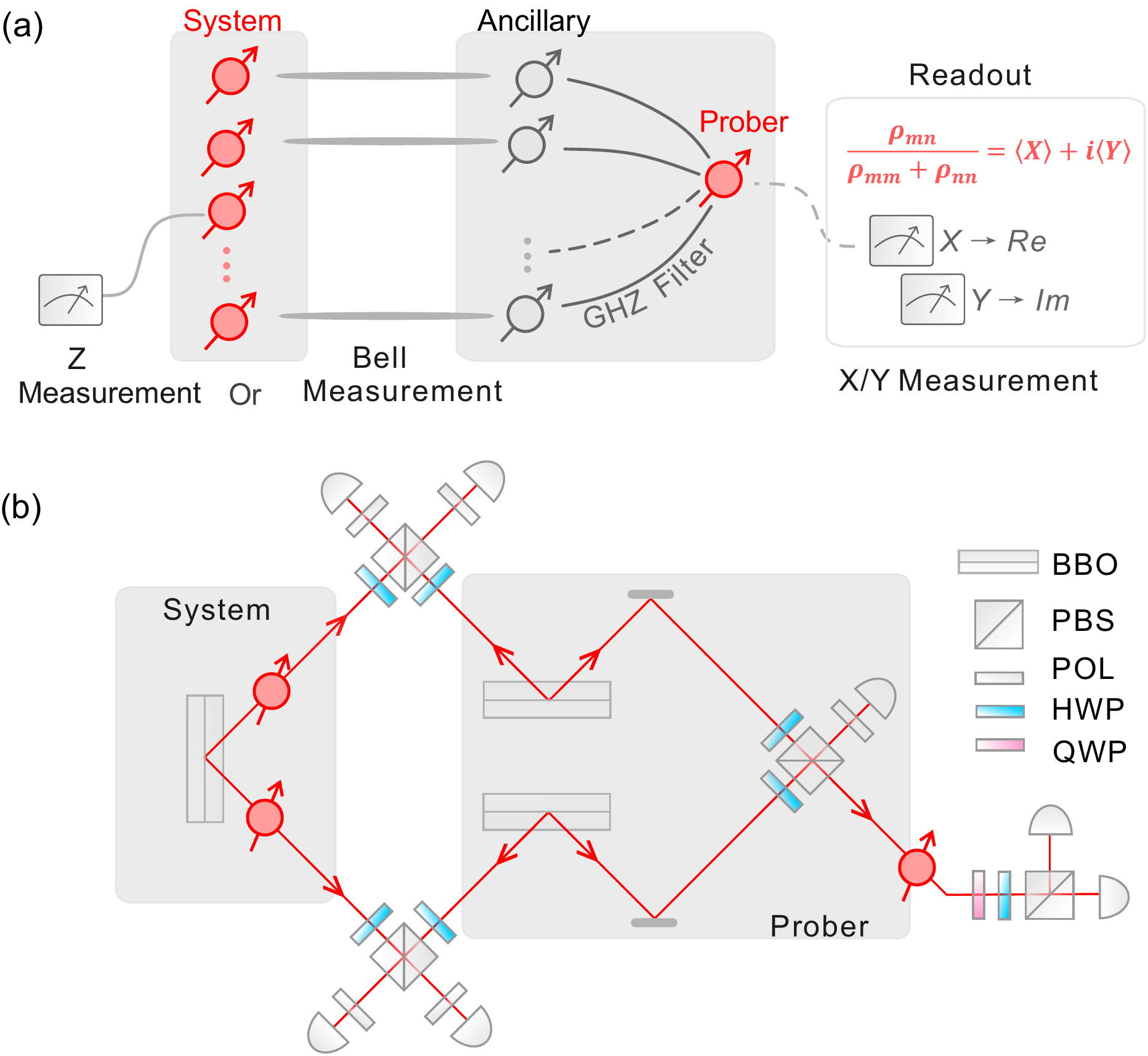}
	\caption{ The protocol. (a) Logical-qubit teleportation for $\rho_{mn}$. The quantum prober is first entangled with the ancillary qubits in GHZ state. The $i$th system qubit is sent to one of two measurement setups ($Z$ or $Bell$ measurements) according to the parity of $m_i$ and $n_i$. The real and imaginary parts of the matrix element are directly obtained by $X$ and $Y$ observables on the prober, respectively. The diagram here shows the case of $m_1 \neq n_1$, $m_2 \neq n_2$, $m_3 = n_3$ and $m_N \neq n_N$. (b) Experimental setup for direct measurement of a general two-photon quantum state. The qubit is encoded in  horizontal (H) and vertical (V) polarization of single photons. Three pairs of EPR polarization-entangled photons are produced by three double-BBO crystals. One pair of entangled photons is used to prepare the two-photon state. Two pairs of entangled photons are used to produce the prober-ancillary GHZ state by two-photon interference on a PBS. The photonic $Bell$ measurement is performed by overlapping two photons on a PBS and erasing the which-way information with two subsequent polarizers. The single photons are detected by single-photon detectors and six-fold coincidence events are recorded. (BBO, beta-barium borate crystal. PBS, polarizing beamsplitter. HWP, half-wave plate. QWP, quarter-wave plate. POL, polarizer)
}
	\label{fig2}
\end{figure}

\begin{figure}[tb]
	\centering
	\includegraphics[width=0.4\textwidth]{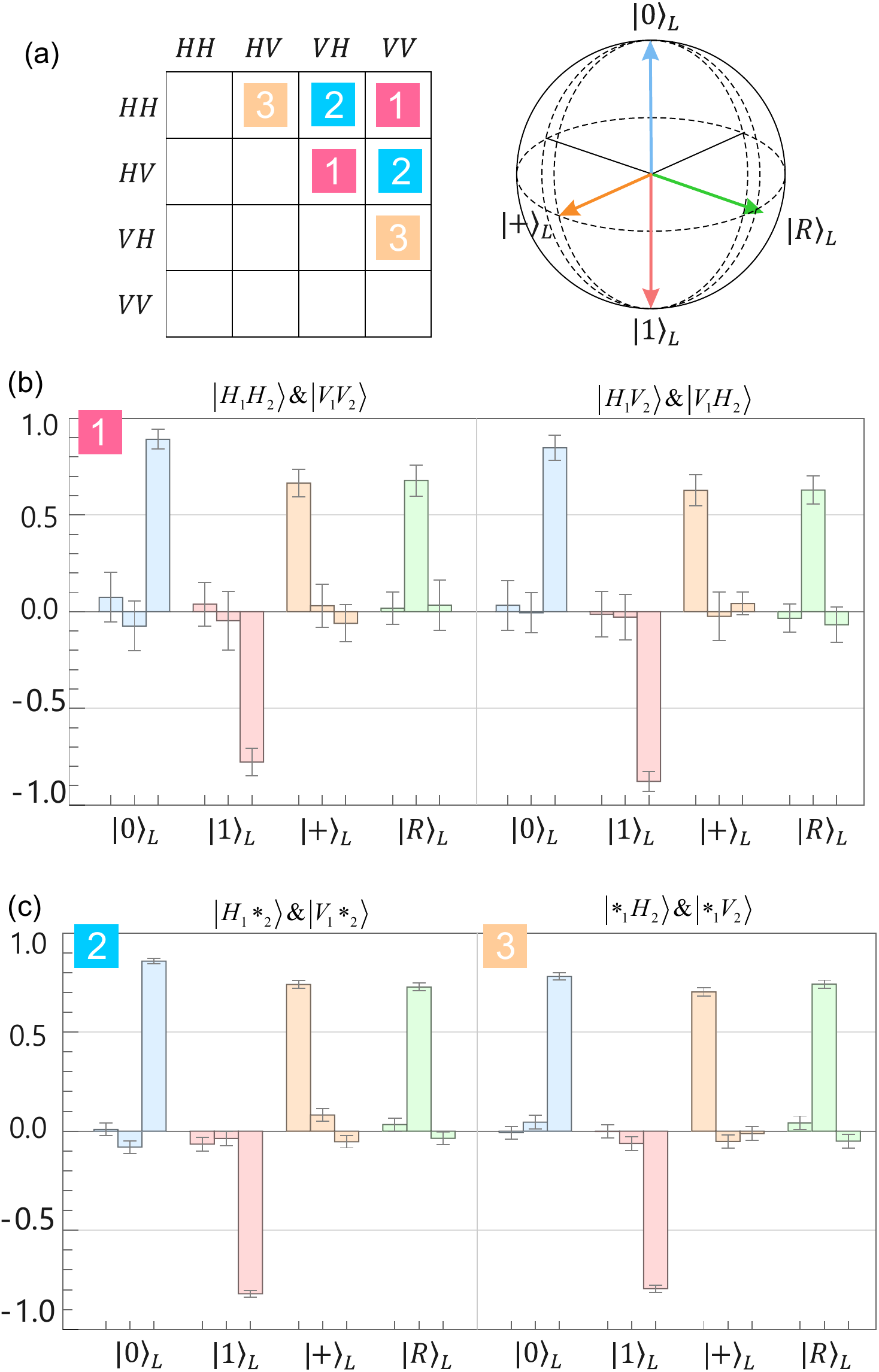}
	\caption{Characterization of the logical-qubit teleporters.  (a) There are $6\times 2$ off-diagonal matrix elements in a two-qubit quantum state, which are embedded in 3 distinct classes of logical qubits according to the parities of the element’s indices.
	Each teleporter is tested by 4 unbiased input states  ${\left| 0 \right\rangle _L}$, ${\left| 1 \right\rangle _L}$,  ${\left| + \right\rangle _L}$, and ${\left| R \right\rangle _L}$.
	(b) Experimental results of teleporter 1, which are obtained by six-photon coincidence counting. Each tested state is measured in $X$, $Y$, and $Z$ bases, and the three expectations with error bars of one standard deviation are shown in same-color groups. (c) Experimental results of the teleporter 2 and 3 by four-photon coincidence counting. The average fidelity of all the teleported states is 0.880(21). }
	\label{fig4}
\end{figure}

In this work, we introduce a simple strategy for direct measurement of multi-particle quantum states using a single observable and without complicated coherent dynamics. 
This method uses quantum teleportation \cite{bennett1993teleporting} to directly address and extract an individual density matrix element to a single physical qubit. It reduces a hard multi-particle measurement problem to an easy single-particle problem.
Next, we describe how the method works.

Considering the simplest quantum system, a single qubit with density matrix $\left( {\begin{array}{*{20}{c}}
{{\rho _{00}}}&{{\rho _{01}}}\\
{{\rho _{10}}}&{{\rho _{11}}}
\end{array}} \right)$, 
the task of direct measurement is naturally achieved. 
The non-negative diagonal matrix elements $\rho _{00}$ and $\rho_{11}$ can be obtained by $Z$-basis measurement.
The real and imaginary parts of complex off-diagonal matrix elements $\rho_{01}$ and $\rho_{10}$ can be obtained by $X$ and $Y$-basis measurement, respectively. 
When scaling the system to $N$ qubits, the diagonal elements (population distribution) can still be accessed by ${Z^{ \otimes N}}$ measurement.
However, the direct measurement of off-diagonal elements become infeasible due to the presence of many off-diagonal elements at the same time. 
To overcome this obstacle, the key insight is that a desired off-diagonal element $\rho_{mn}$ is indeed embedded in a single virtual logical qubit $\left( {\begin{array}{*{20}{c}}
{{\rho _{mm}}}&{{\rho _{mn}}}\\
{{\rho _{nm}}}&{{\rho _{nn}}}
\end{array}} \right)$. 
If we are able to nondestructively transfer the quantum information of this logical qubit to a single physical qubit (see Fig.\ \ref{fig1}), the problem will be \emph{reduced} to the aforementioned single-qubit measurement problem.

The faithful transfer of a quantum state from the virtual logical qubit to a physical qubit can be realized in the framework of quantum teleportation.
Quantum teleportation uses quantum entanglement to transport quantum information, which has played important roles in various quantum information technologies, such as distributed quantum networks \cite{kimble2008quantum} and measurement-based quantum computation \cite{gottesman1999demonstrating, raussendorf2001one}. 
For our current purpose, quantum teleportation serves dual roles:  acting as a quantum state filter \cite{okamoto2009entanglement} and acting as a quantum state channel \cite{bennett1993teleporting}. 
The quantum state filter will block the undesired off-diagonal elements except for the logical qubit.
Meanwhile, the logical qubit will be transported to a physical qubit, i.e. the quantum prober, through the quantum teleportation channel. 
Next, we show how to build such a teleporter.

\textbf{Logical-qubit Teleportation.}
The density matrix element ${\rho _{mn}}$ in $N$-qubit quantum state indicates the coherence between components $|m\rangle $ and $|n\rangle $, where the indices $m$ and $n$ can be expressed in binary representation as ${m_1}{m_2} \cdots {m_i} \cdots {m_N}$ and ${n_1}{n_2} \cdots {n_i} \cdots {n_N}$ with ${m_i},{n_i} \in \{ 0,1\} $.  
This individual element ${\rho _{mn}}$, together with three relevant elements ${\rho _{mm}}$, ${\rho _{nn}}$, and ${\rho _{nm}}$ compose a distributed logical qubit, as shown in Fig.\ \ref{fig1}. 
To filter out and teleport the logical qubit to an external physical qubit, the $i$th system qubit is sent to one of the two following measurement setups according to the \emph{bit-wise parity} of $m_{i}$ and $n_{i}$, as shown in Fig.\ \ref{fig2}(a).  
If $m_{i}=n_{i}$, the $i$th system qubit is directly measured in the $Z$ basis. 
If ${m_i} \ne {n_i}$, the $i$th system qubit is jointly measured with an ancillary qubit in the $Bell$ basis. 
These ancillary qubits are initially entangled with a single-qubit prober in Greenberger-Horne-Zeilinger (GHZ) state \cite{greenberger1990bell}. The size of GHZ state is $k+1$ qubits, where $k (\le N) $ is the number of bits that have odd parity in the indices $m$ and $n$. 
In the following, we denote the bit sets of system qubits in $Z$ measurement and $Bell$ measurement as $\{i\}$ and $\{j\}$, respectively.

When the results of $Z$ measurements on the system qubits $\{i\}$ are $\{ \left| m_i \right\rangle \}$ and 
the results of $Bell$ measurements on system qubits $\{j\}$ are $\{ X_1^{m_j}(  \left| {00} \right\rangle  + \left| {11} \right\rangle )/\sqrt 2 \}$, respectively,
the logical qubit $\alpha \left| m \right\rangle  + \beta \left| n \right\rangle $ is successfully teleported to the quantum prober. The prober is in a quantum state $\alpha \left| 0 \right\rangle  + \beta \left| 1 \right\rangle $ up to normalization.
The linearity of quantum process will ensure that it is also applicable to a logical qubit with mixed states.

After successful teleportion, the desired density matrix element ${\rho _{mn}}$ is embedded in the quantum prober and its real and imaginary parts can be read out directly by $X$ and $Y$ measurements, according to the following equation \begin{equation}  {\tilde \rho _{mn}} = \frac{{{\rho _{mn}}}}{{{\rho _{mm}} + {\rho _{nn}}}} = \langle X\rangle  + i\langle Y\rangle. \end{equation}

In general, there are four different results in each $Bell$ measurement and two different results in each $Z$ measurement. These possible results corresponding to different logical qubits are addressed and teleported, which is a useful feature to scan all the nonzero density matrix elements for full quantum state characterization. 
Concretely,  the teleported logical qubit is in the 2-dimension subspace $\{ \hat P\left| m \right\rangle ,\hat P\left| n \right\rangle \} $, where $\hat P$ is an $N$-body Pauli correction operator determined by the random results of the $Bell$ and $Z$ measurements. The probability of teleporting out the information in subspace  $\{ \left| m \right\rangle , \left| n \right\rangle \} $  is  ${\rho _{mm}} + {\rho _{nn}}$, which indicates that the off-diagonal elements are sampled according to the importance weighting in the population distribution.

\textbf{Experimental demonstration.} 
The quantum resource used in logical-qubit teleportation is GHZ entangled states and $Bell$ measurements,  which are standard routines of quantum optics experiments and have been widely realized in photonic \cite{bouwmeester1997experimental, zhao2004experimental, zhang2006experimental, wang2015quantum}, atomic  \cite{krauter2013deterministic, riebe2004deterministic, barrett2004deterministic}, and solid system \cite{pfaff2014unconditional, steffen2013deterministic}. Next, we demonstrate the protocol to directly measure a general two-photon quantum state.

The experimental setup is shown in Fig.\ \ref{fig2}(b). This measurement configuration is set to teleport out the elements $\rho_{14}$ or $\rho_{23}$ in the case of ${m_1} \ne {n_1}$ and ${m_2} \ne {n_2}$. The qubits  are encoded in the horizontal (H) and vertical (V) polarization of a single photon. Three Einstein-Podolsky-Rosen (EPR) polarization-entangled photon pairs are produced by beam-like type-II pulsed SPDC processes from three double-BBO crystals \cite{zhang2015experimental, 10photons} pumping by  an Ultraviolet femtosecond laser (394nm, 120fs, 80MHz). The average fidelity of EPR photon pairs is $0.959$ under $400$ mW pumping power with narrowband filters. The first EPR photon pair is used to engineer the two-photon quantum state by local operations. The second and third photon pairs are used to produce GHZ entangled state by mixing two single photons from each pair on a polarizing beamsplitter (PBS). We trace out an extra photon by projecting it onto diagonal polarization and obtain the three-photon ancillary-prober GHZ state. The system-ancillary $Bell$ measurements are implemented by overlapping two photons on a PBS and subsequently erase the which-way information by a diagonal polarizer at each output port. Here, we use post-selection to  distinguish single Bell basis and in general, feed-forward of the results of Bell measurement can be used to improve the efficiency of teleportation. Finally, the quantum state of prober is measured in $X$ or $Y$ basis using quarter-wave
plates (QWPs), half-wave plates (HWPs) and PBSs. In the photonic interferometer, only the events where one and only one single photon in each optical path mode are selected. All photons are detected by fiber-coupling silicon avalanched single-photon detectors and six-fold coincidence events are recorded.

\begin{figure}[tb]
	\centering
	\includegraphics[width=0.4\textwidth]{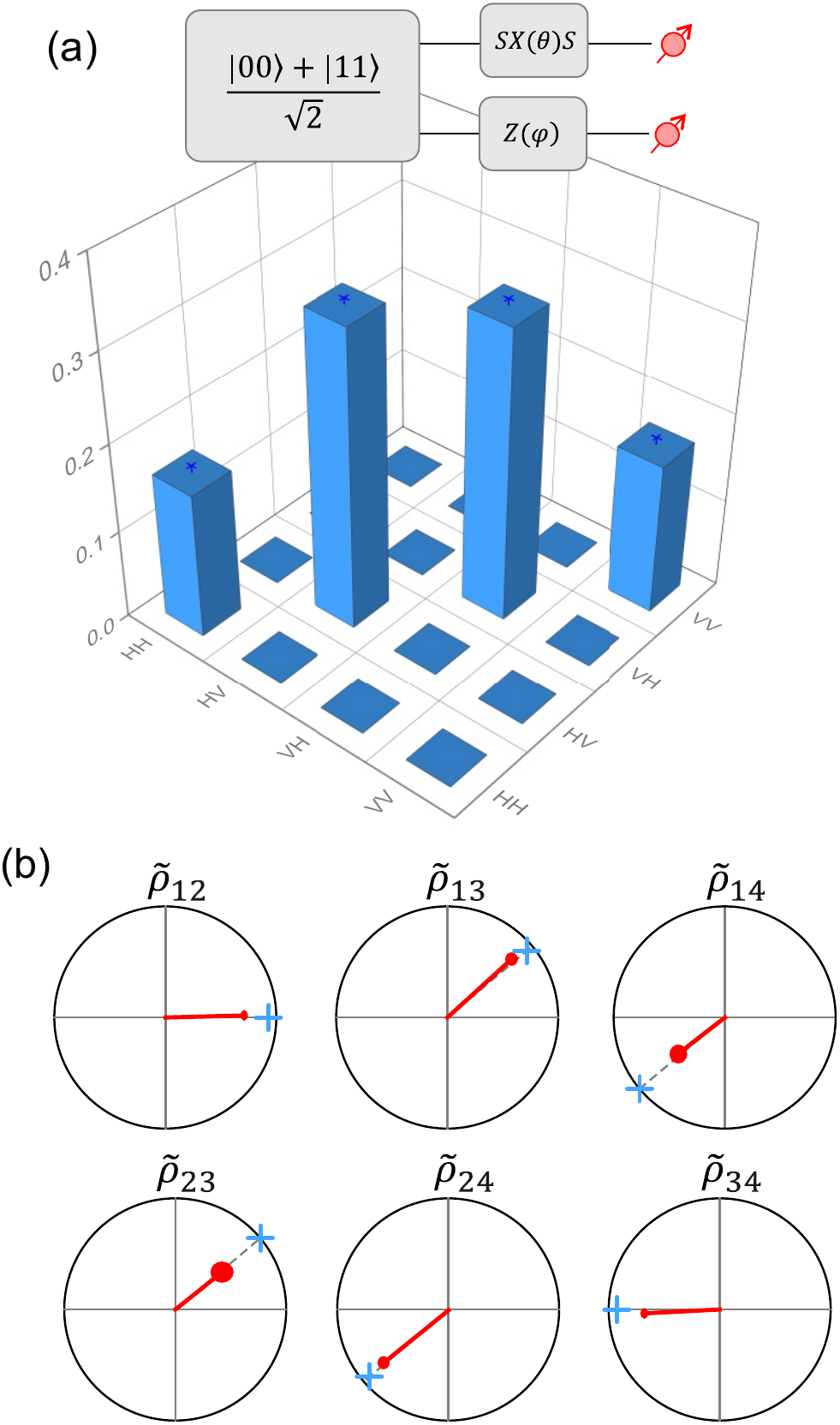}
	\caption{Efficient two-step measurement strategy. An instance of general two-photon quantum states is generated by locally rotating an EPR state with parameters $\theta  = {56^ \circ }$ and $\varphi  = {20^ \circ }$ in the quantum gates. (a) The population distribution is obtained by ${Z^{ \otimes 2}}$ measurement at the first step and showed on the diagonal line of a density matrix. (b) The complex-valued d off-diagonal element $\rho_{mn}$ is measured at $X$ and $Y$ bases at the second step, if the relevant populations $\rho_{mm}$ and $\rho_{nn}$ are significantly non-zero. The results are shown as complex vectors on complex planes. One standard deviation of the real and imaginary parts is represented by the elliptic data points. The blue symbols $+$ indicate the ideal values. }
	\label{fig5}
\end{figure}

The experiment is carried out in two stages. In the first stage, the capability of the quantum teleportation channel to preserve the quantum information of logical qubits is verified. We benchmark the performance of the teleporter by feeding it with a set of tested quantum states and measuring the fidelity of the outputs. 
For two-qubit quantum states, there are three distinct classes of the off-diagonal matrix elements (labeled as $1\sim3$ in Fig.\ \ref{fig4}(a), respectively), according to the parities of the off-diagonal element's indices. We use four tested quantum states ${\left| 0 \right\rangle _L}$, ${\left| 1 \right\rangle _L}$,  ${\left| + \right\rangle _L}$, and ${\left| R \right\rangle _L}$ to evaluate each teleporter. 
We show the measured results of $X$,$Y$, and $Z$ observables of the four tested states for each teleporter in Fig.\ \ref{fig4}(b) and \ref{fig4}(c).
We use these results to reconstruct the channel matrices and obtain quantum process fidelities of $0.800$, $0.782$, $0.848$, and $0.811$ for the four teleporters, respectively.

In the second stage, we test the teleporter with general quantum states to prove its ability to address a single logical qubit while blocking the other components.
We prepared an instance of tested quantum state by locally rotating an EPR state, which has four non-zero components. 
We first measure the population distribution and show it in Fig.\ \ref{fig5}(a).
We use the teleporters configured in the first stage to extract each individual off-diagonal element to a quantum prober. Then, the real and imaginary parts of the off-diagonal elements are read out by $X$ and $Y$ measurements on the prober, respectively.
The experimental results are shown in Fig.\ \ref{fig5}(b), where the complex-valued d matrix elements are visualized as vectors on complex planes.
One standard deviation of the real and imaginary parts is represented by the size of elliptic data points. 

The phase information of the measured off-diagonal elements agrees well with the ideal values (with a mean deviation of $0.010\pi $), due to the automatic phase calibration in our interferometer. The measured amplitudes are shrunk from the ideal values by scaling factors $0.811$, $0.844$, $0.538$, $0.533$, $0.766$ and $0.733$, respectively. The reduction in the amplitude visibility is mainly caused by the finite purities of six-photon states. The purity can be improved by engineering an optimized quantum light source without spectral correlation in the SPDC process.

Our results provide an efficient method for measuring a full quantum state. For a general $N$-qubit quantum state, there are ~$4^N$ independent parameters and it needs ~$3^N$ measurement settings to characterize the state. However, if a quantum state has certain  mathematical structures, the independent parameters and thus the measurement settings will reduce significantly. For example, quantum state tomography based on quantum state ansatzes, e.g. matrix product states \cite{cramer2010efficient}, neural network quantum states \cite{torlai2018neural}, and compressed sensing for low-rank density matrix have been proposed \cite{gross2010quantum}. Here, we note that if a quantum state is sparse in the population distribution, our method of logical-qubit teleportation can exponentially reduce the measurement settings comparing to the standard quantum state tomography.

This is due to an off-diagonal element $\rho_{mn}$ is non-zero only when the relevant diagonal elements $\rho_{mm}$  and $\rho_{nn}$ are significantly non-zero. This constraint $\left| {{\rho _{mn}}} \right| \le \sqrt {{\rho _{mm}}{\rho _{nn}}} $ suggests a two-step measurement strategy for sparse quantum states: (1) measure the population distribution; (2) measure the potential non-zero off-diagonal elements. 
A sparse quantum state with $s (\sim poly(N))$ non-zero diagonal elements has at most $s(s - 1)$ non-zero off-diagonal elements. Therefore, it takes $\sim s^2$ measurement settings to directly measure the full state and avoids the complicated post-processing of the measured data. We further note that our method can mitigate certain noise in the ancillary-prober state. For example, with a noisy GHZ state ${\rho } = p\left| {GHZ} \right\rangle \left\langle {GHZ} \right| + (1 - p){I \over {{2^N}}}$, the equation (1) of the measured elements becomes ${{\tilde \rho }_{mn}} = p{{{\rho _{mn}}} \over {{\rho _{mm}} + {\rho _{nn}}}}$. By in advance characterizing the noise level $p$, the true element $\rho _{mm}$ can be recovered.

Our method also provides a number of applications in quantum information protocols. An immediate one is to work as a quantum filter. It can non-destructively extract one-qubit sub-space quantum information embedding in distributed quantum states. Another application is to implement quantum metrology for the relative quantum phase between two components in many-body quantum states, which extends quantum metrology to non-classical phases and can achieve quantum metrology advantage with only small-scale quantum entanglement resources.

In summary, we have introduced and demonstrated a novel and efficient  method to directly measure multi-particle quantum wavefuntion based on the concept of quantum teleportation. Our protocol has many interesting features: it is non-destructive to extract single quantum amplitudes; it uses single observable to get the real or imagine part of complex quantum amplitudes and it is very general that can apply to non-pure states and many-qubit states.

~\\

\textit{Acknowledgement}:  This work was supported by the National Natural Science Foundation of China, the Chinese Academy of Sciences, the National Fundamental Research, the Anhui Initiative in Quantum Information Technologies.

\bibliographystyle{apsrev4-1}
\bibliography{mybibtex}

\begin{thebibliography}{40}%
\makeatletter
\providecommand \@ifxundefined [1]{%
 \@ifx{#1\undefined}
}%
\providecommand \@ifnum [1]{%
 \ifnum #1\expandafter \@firstoftwo
 \else \expandafter \@secondoftwo
 \fi
}%
\providecommand \@ifx [1]{%
 \ifx #1\expandafter \@firstoftwo
 \else \expandafter \@secondoftwo
 \fi
}%
\providecommand \natexlab [1]{#1}%
\providecommand \enquote  [1]{``#1''}%
\providecommand \bibnamefont  [1]{#1}%
\providecommand \bibfnamefont [1]{#1}%
\providecommand \citenamefont [1]{#1}%
\providecommand \href@noop [0]{\@secondoftwo}%
\providecommand \href [0]{\begingroup \@sanitize@url \@href}%
\providecommand \@href[1]{\@@startlink{#1}\@@href}%
\providecommand \@@href[1]{\endgroup#1\@@endlink}%
\providecommand \@sanitize@url [0]{\catcode `\\12\catcode `\$12\catcode
  `\&12\catcode `\#12\catcode `\^12\catcode `\_12\catcode `\%12\relax}%
\providecommand \@@startlink[1]{}%
\providecommand \@@endlink[0]{}%
\providecommand \url  [0]{\begingroup\@sanitize@url \@url }%
\providecommand \@url [1]{\endgroup\@href {#1}{\urlprefix }}%
\providecommand \urlprefix  [0]{URL }%
\providecommand \Eprint [0]{\href }%
\providecommand \doibase [0]{http://dx.doi.org/}%
\providecommand \selectlanguage [0]{\@gobble}%
\providecommand \bibinfo  [0]{\@secondoftwo}%
\providecommand \bibfield  [0]{\@secondoftwo}%
\providecommand \translation [1]{[#1]}%
\providecommand \BibitemOpen [0]{}%
\providecommand \bibitemStop [0]{}%
\providecommand \bibitemNoStop [0]{.\EOS\space}%
\providecommand \EOS [0]{\spacefactor3000\relax}%
\providecommand \BibitemShut  [1]{\csname bibitem#1\endcsname}%
\let\auto@bib@innerbib\@empty
\bibitem [{\citenamefont {Dirac}(1981)}]{dirac1981}%
  \BibitemOpen
  \bibfield  {author} {\bibinfo {author} {\bibfnamefont {P.~A.~M.}\
  \bibnamefont {Dirac}},\ }\href@noop {} {\emph {\bibinfo {title} {The
  principles of quantum mechanics}}},\ \bibinfo {number} {27}\ (\bibinfo
  {publisher} {Oxford university press},\ \bibinfo {year} {1981})\BibitemShut
  {NoStop}%
\bibitem [{\citenamefont {Nielsen}\ and\ \citenamefont
  {Chuang}(2010)}]{nielsen2010}%
  \BibitemOpen
  \bibfield  {author} {\bibinfo {author} {\bibfnamefont {M.~A.}\ \bibnamefont
  {Nielsen}}\ and\ \bibinfo {author} {\bibfnamefont {I.~L.}\ \bibnamefont
  {Chuang}},\ }\href@noop {} {\emph {\bibinfo {title} {Quantum Computation and
  Quantum Information}}}\ (\bibinfo  {publisher} {Cambridge University Press},\
  \bibinfo {year} {2010})\BibitemShut {NoStop}%
\bibitem [{\citenamefont {Giovannetti}\ \emph {et~al.}(2011)\citenamefont
  {Giovannetti}, \citenamefont {Lloyd},\ and\ \citenamefont
  {Maccone}}]{metrology}%
  \BibitemOpen
  \bibfield  {author} {\bibinfo {author} {\bibfnamefont {V.}~\bibnamefont
  {Giovannetti}}, \bibinfo {author} {\bibfnamefont {S.}~\bibnamefont {Lloyd}},
  \ and\ \bibinfo {author} {\bibfnamefont {L.}~\bibnamefont {Maccone}},\
  }\href@noop {} {\bibfield  {journal} {\bibinfo  {journal} {Nature photonics}\
  }\textbf {\bibinfo {volume} {5}},\ \bibinfo {pages} {222} (\bibinfo {year}
  {2011})}\BibitemShut {NoStop}%
\bibitem [{\citenamefont {Born}(1955)}]{born1955}%
  \BibitemOpen
  \bibfield  {author} {\bibinfo {author} {\bibfnamefont {M.}~\bibnamefont
  {Born}},\ }\href@noop {} {\bibfield  {journal} {\bibinfo  {journal}
  {Science}\ }\textbf {\bibinfo {volume} {122}},\ \bibinfo {pages} {675}
  (\bibinfo {year} {1955})}\BibitemShut {NoStop}%
\bibitem [{\citenamefont {Pusey}\ \emph {et~al.}(2012)\citenamefont {Pusey},
  \citenamefont {Barrett},\ and\ \citenamefont {Rudolph}}]{pusey2012reality}%
  \BibitemOpen
  \bibfield  {author} {\bibinfo {author} {\bibfnamefont {M.~F.}\ \bibnamefont
  {Pusey}}, \bibinfo {author} {\bibfnamefont {J.}~\bibnamefont {Barrett}}, \
  and\ \bibinfo {author} {\bibfnamefont {T.}~\bibnamefont {Rudolph}},\
  }\href@noop {} {\bibfield  {journal} {\bibinfo  {journal} {Nature Physics}\
  }\textbf {\bibinfo {volume} {8}},\ \bibinfo {pages} {475} (\bibinfo {year}
  {2012})}\BibitemShut {NoStop}%
\bibitem [{\citenamefont {Hradil}(1997)}]{hradil1997tomo}%
  \BibitemOpen
  \bibfield  {author} {\bibinfo {author} {\bibfnamefont {Z.}~\bibnamefont
  {Hradil}},\ }\href@noop {} {\bibfield  {journal} {\bibinfo  {journal}
  {Physical Review A}\ }\textbf {\bibinfo {volume} {55}},\ \bibinfo {pages}
  {R1561} (\bibinfo {year} {1997})}\BibitemShut {NoStop}%
\bibitem [{\citenamefont {Wang}\ \emph {et~al.}(2016)\citenamefont {Wang},
  \citenamefont {Chen}, \citenamefont {Li}, \citenamefont {Huang},
  \citenamefont {Liu}, \citenamefont {Chen}, \citenamefont {Luo}, \citenamefont
  {Su}, \citenamefont {Wu}, \citenamefont {Li} \emph {et~al.}}]{10photons}%
  \BibitemOpen
  \bibfield  {author} {\bibinfo {author} {\bibfnamefont {X.-L.}\ \bibnamefont
  {Wang}}, \bibinfo {author} {\bibfnamefont {L.-K.}\ \bibnamefont {Chen}},
  \bibinfo {author} {\bibfnamefont {W.}~\bibnamefont {Li}}, \bibinfo {author}
  {\bibfnamefont {H.-L.}\ \bibnamefont {Huang}}, \bibinfo {author}
  {\bibfnamefont {C.}~\bibnamefont {Liu}}, \bibinfo {author} {\bibfnamefont
  {C.}~\bibnamefont {Chen}}, \bibinfo {author} {\bibfnamefont {Y.-H.}\
  \bibnamefont {Luo}}, \bibinfo {author} {\bibfnamefont {Z.-E.}\ \bibnamefont
  {Su}}, \bibinfo {author} {\bibfnamefont {D.}~\bibnamefont {Wu}}, \bibinfo
  {author} {\bibfnamefont {Z.-D.}\ \bibnamefont {Li}},  \emph {et~al.},\
  }\href@noop {} {\bibfield  {journal} {\bibinfo  {journal} {Physical review
  letters}\ }\textbf {\bibinfo {volume} {117}},\ \bibinfo {pages} {210502}
  (\bibinfo {year} {2016})}\BibitemShut {NoStop}%
\bibitem [{\citenamefont {Monz}\ \emph {et~al.}(2011)\citenamefont {Monz},
  \citenamefont {Schindler}, \citenamefont {Barreiro}, \citenamefont {Chwalla},
  \citenamefont {Nigg}, \citenamefont {Coish}, \citenamefont {Harlander},
  \citenamefont {H{\"a}nsel}, \citenamefont {Hennrich},\ and\ \citenamefont
  {Blatt}}]{14ions}%
  \BibitemOpen
  \bibfield  {author} {\bibinfo {author} {\bibfnamefont {T.}~\bibnamefont
  {Monz}}, \bibinfo {author} {\bibfnamefont {P.}~\bibnamefont {Schindler}},
  \bibinfo {author} {\bibfnamefont {J.~T.}\ \bibnamefont {Barreiro}}, \bibinfo
  {author} {\bibfnamefont {M.}~\bibnamefont {Chwalla}}, \bibinfo {author}
  {\bibfnamefont {D.}~\bibnamefont {Nigg}}, \bibinfo {author} {\bibfnamefont
  {W.~A.}\ \bibnamefont {Coish}}, \bibinfo {author} {\bibfnamefont
  {M.}~\bibnamefont {Harlander}}, \bibinfo {author} {\bibfnamefont
  {W.}~\bibnamefont {H{\"a}nsel}}, \bibinfo {author} {\bibfnamefont
  {M.}~\bibnamefont {Hennrich}}, \ and\ \bibinfo {author} {\bibfnamefont
  {R.}~\bibnamefont {Blatt}},\ }\href@noop {} {\bibfield  {journal} {\bibinfo
  {journal} {Physical Review Letters}\ }\textbf {\bibinfo {volume} {106}},\
  \bibinfo {pages} {130506} (\bibinfo {year} {2011})}\BibitemShut {NoStop}%
\bibitem [{\citenamefont {Song}\ \emph {et~al.}(2017)\citenamefont {Song},
  \citenamefont {Xu}, \citenamefont {Liu}, \citenamefont {Yang}, \citenamefont
  {Zheng}, \citenamefont {Deng}, \citenamefont {Xie}, \citenamefont {Huang},
  \citenamefont {Guo}, \citenamefont {Zhang} \emph {et~al.}}]{10Xmons}%
  \BibitemOpen
  \bibfield  {author} {\bibinfo {author} {\bibfnamefont {C.}~\bibnamefont
  {Song}}, \bibinfo {author} {\bibfnamefont {K.}~\bibnamefont {Xu}}, \bibinfo
  {author} {\bibfnamefont {W.}~\bibnamefont {Liu}}, \bibinfo {author}
  {\bibfnamefont {C.}~\bibnamefont {Yang}}, \bibinfo {author} {\bibfnamefont
  {S.-B.}\ \bibnamefont {Zheng}}, \bibinfo {author} {\bibfnamefont
  {H.}~\bibnamefont {Deng}}, \bibinfo {author} {\bibfnamefont {Q.}~\bibnamefont
  {Xie}}, \bibinfo {author} {\bibfnamefont {K.}~\bibnamefont {Huang}}, \bibinfo
  {author} {\bibfnamefont {Q.}~\bibnamefont {Guo}}, \bibinfo {author}
  {\bibfnamefont {L.}~\bibnamefont {Zhang}},  \emph {et~al.},\ }\href@noop {}
  {\bibfield  {journal} {\bibinfo  {journal} {arXiv preprint arXiv:1703.10302}\
  } (\bibinfo {year} {2017})}\BibitemShut {NoStop}%
\bibitem [{\citenamefont {Lundeen}\ \emph {et~al.}(2011)\citenamefont
  {Lundeen}, \citenamefont {Sutherland}, \citenamefont {Patel}, \citenamefont
  {Stewart},\ and\ \citenamefont {Bamber}}]{lundeen2011direct}%
  \BibitemOpen
  \bibfield  {author} {\bibinfo {author} {\bibfnamefont {J.~S.}\ \bibnamefont
  {Lundeen}}, \bibinfo {author} {\bibfnamefont {B.}~\bibnamefont {Sutherland}},
  \bibinfo {author} {\bibfnamefont {A.}~\bibnamefont {Patel}}, \bibinfo
  {author} {\bibfnamefont {C.}~\bibnamefont {Stewart}}, \ and\ \bibinfo
  {author} {\bibfnamefont {C.}~\bibnamefont {Bamber}},\ }\href@noop {}
  {\bibfield  {journal} {\bibinfo  {journal} {Nature}\ }\textbf {\bibinfo
  {volume} {474}},\ \bibinfo {pages} {188} (\bibinfo {year}
  {2011})}\BibitemShut {NoStop}%
\bibitem [{\citenamefont {Lundeen}\ and\ \citenamefont
  {Bamber}(2012)}]{lundeen2012procedure}%
  \BibitemOpen
  \bibfield  {author} {\bibinfo {author} {\bibfnamefont {J.~S.}\ \bibnamefont
  {Lundeen}}\ and\ \bibinfo {author} {\bibfnamefont {C.}~\bibnamefont
  {Bamber}},\ }\href@noop {} {\bibfield  {journal} {\bibinfo  {journal}
  {Physical review letters}\ }\textbf {\bibinfo {volume} {108}},\ \bibinfo
  {pages} {070402} (\bibinfo {year} {2012})}\BibitemShut {NoStop}%
\bibitem [{\citenamefont {Wu}(2013)}]{wu2013state}%
  \BibitemOpen
  \bibfield  {author} {\bibinfo {author} {\bibfnamefont {S.}~\bibnamefont
  {Wu}},\ }\href@noop {} {\bibfield  {journal} {\bibinfo  {journal} {Scientific
  reports}\ }\textbf {\bibinfo {volume} {3}},\ \bibinfo {pages} {1193}
  (\bibinfo {year} {2013})}\BibitemShut {NoStop}%
\bibitem [{\citenamefont {Vallone}\ and\ \citenamefont
  {Dequal}(2016)}]{vallone2016strong}%
  \BibitemOpen
  \bibfield  {author} {\bibinfo {author} {\bibfnamefont {G.}~\bibnamefont
  {Vallone}}\ and\ \bibinfo {author} {\bibfnamefont {D.}~\bibnamefont
  {Dequal}},\ }\href@noop {} {\bibfield  {journal} {\bibinfo  {journal}
  {Physical review letters}\ }\textbf {\bibinfo {volume} {116}},\ \bibinfo
  {pages} {040502} (\bibinfo {year} {2016})}\BibitemShut {NoStop}%
\bibitem [{\citenamefont {Zou}\ \emph {et~al.}(2015)\citenamefont {Zou},
  \citenamefont {Zhang},\ and\ \citenamefont {Song}}]{zou2015direct}%
  \BibitemOpen
  \bibfield  {author} {\bibinfo {author} {\bibfnamefont {P.}~\bibnamefont
  {Zou}}, \bibinfo {author} {\bibfnamefont {Z.-M.}\ \bibnamefont {Zhang}}, \
  and\ \bibinfo {author} {\bibfnamefont {W.}~\bibnamefont {Song}},\ }\href@noop
  {} {\bibfield  {journal} {\bibinfo  {journal} {Physical Review A}\ }\textbf
  {\bibinfo {volume} {91}},\ \bibinfo {pages} {052109} (\bibinfo {year}
  {2015})}\BibitemShut {NoStop}%
\bibitem [{\citenamefont {Malik}\ \emph {et~al.}(2014)\citenamefont {Malik},
  \citenamefont {Mirhosseini}, \citenamefont {Lavery}, \citenamefont {Leach},
  \citenamefont {Padgett},\ and\ \citenamefont {Boyd}}]{malik2014direct}%
  \BibitemOpen
  \bibfield  {author} {\bibinfo {author} {\bibfnamefont {M.}~\bibnamefont
  {Malik}}, \bibinfo {author} {\bibfnamefont {M.}~\bibnamefont {Mirhosseini}},
  \bibinfo {author} {\bibfnamefont {M.~P.}\ \bibnamefont {Lavery}}, \bibinfo
  {author} {\bibfnamefont {J.}~\bibnamefont {Leach}}, \bibinfo {author}
  {\bibfnamefont {M.~J.}\ \bibnamefont {Padgett}}, \ and\ \bibinfo {author}
  {\bibfnamefont {R.~W.}\ \bibnamefont {Boyd}},\ }\href@noop {} {\bibfield
  {journal} {\bibinfo  {journal} {Nature Communications}\ }\textbf {\bibinfo
  {volume} {5}},\ \bibinfo {pages} {3115} (\bibinfo {year} {2014})}\BibitemShut
  {NoStop}%
\bibitem [{\citenamefont {Shi}\ \emph {et~al.}(2015)\citenamefont {Shi},
  \citenamefont {Mirhosseini}, \citenamefont {Margiewicz}, \citenamefont
  {Malik}, \citenamefont {Rivera}, \citenamefont {Zhu},\ and\ \citenamefont
  {Boyd}}]{shi2015scan}%
  \BibitemOpen
  \bibfield  {author} {\bibinfo {author} {\bibfnamefont {Z.}~\bibnamefont
  {Shi}}, \bibinfo {author} {\bibfnamefont {M.}~\bibnamefont {Mirhosseini}},
  \bibinfo {author} {\bibfnamefont {J.}~\bibnamefont {Margiewicz}}, \bibinfo
  {author} {\bibfnamefont {M.}~\bibnamefont {Malik}}, \bibinfo {author}
  {\bibfnamefont {F.}~\bibnamefont {Rivera}}, \bibinfo {author} {\bibfnamefont
  {Z.}~\bibnamefont {Zhu}}, \ and\ \bibinfo {author} {\bibfnamefont {R.~W.}\
  \bibnamefont {Boyd}},\ }\href@noop {} {\bibfield  {journal} {\bibinfo
  {journal} {Optica}\ }\textbf {\bibinfo {volume} {2}},\ \bibinfo {pages} {388}
  (\bibinfo {year} {2015})}\BibitemShut {NoStop}%
\bibitem [{\citenamefont {Salvail}\ \emph {et~al.}(2013)\citenamefont
  {Salvail}, \citenamefont {Agnew}, \citenamefont {Johnson}, \citenamefont
  {Bolduc}, \citenamefont {Leach},\ and\ \citenamefont
  {Boyd}}]{salvail2013full}%
  \BibitemOpen
  \bibfield  {author} {\bibinfo {author} {\bibfnamefont {J.~Z.}\ \bibnamefont
  {Salvail}}, \bibinfo {author} {\bibfnamefont {M.}~\bibnamefont {Agnew}},
  \bibinfo {author} {\bibfnamefont {A.~S.}\ \bibnamefont {Johnson}}, \bibinfo
  {author} {\bibfnamefont {E.}~\bibnamefont {Bolduc}}, \bibinfo {author}
  {\bibfnamefont {J.}~\bibnamefont {Leach}}, \ and\ \bibinfo {author}
  {\bibfnamefont {R.~W.}\ \bibnamefont {Boyd}},\ }\href@noop {} {\bibfield
  {journal} {\bibinfo  {journal} {Nature Photonics}\ }\textbf {\bibinfo
  {volume} {7}},\ \bibinfo {pages} {316} (\bibinfo {year} {2013})}\BibitemShut
  {NoStop}%
\bibitem [{\citenamefont {Thekkadath}\ \emph {et~al.}(2016)\citenamefont
  {Thekkadath}, \citenamefont {Giner}, \citenamefont {Chalich}, \citenamefont
  {Horton}, \citenamefont {Banker},\ and\ \citenamefont
  {Lundeen}}]{thekkadath2016direct}%
  \BibitemOpen
  \bibfield  {author} {\bibinfo {author} {\bibfnamefont {G.}~\bibnamefont
  {Thekkadath}}, \bibinfo {author} {\bibfnamefont {L.}~\bibnamefont {Giner}},
  \bibinfo {author} {\bibfnamefont {Y.}~\bibnamefont {Chalich}}, \bibinfo
  {author} {\bibfnamefont {M.}~\bibnamefont {Horton}}, \bibinfo {author}
  {\bibfnamefont {J.}~\bibnamefont {Banker}}, \ and\ \bibinfo {author}
  {\bibfnamefont {J.}~\bibnamefont {Lundeen}},\ }\href@noop {} {\bibfield
  {journal} {\bibinfo  {journal} {Physical review letters}\ }\textbf {\bibinfo
  {volume} {117}},\ \bibinfo {pages} {120401} (\bibinfo {year}
  {2016})}\BibitemShut {NoStop}%
\bibitem [{\citenamefont {Calderaro}\ \emph {et~al.}(2018)\citenamefont
  {Calderaro}, \citenamefont {Foletto}, \citenamefont {Dequal}, \citenamefont
  {Villoresi},\ and\ \citenamefont {Vallone}}]{directMea2018}%
  \BibitemOpen
  \bibfield  {author} {\bibinfo {author} {\bibfnamefont {L.}~\bibnamefont
  {Calderaro}}, \bibinfo {author} {\bibfnamefont {G.}~\bibnamefont {Foletto}},
  \bibinfo {author} {\bibfnamefont {D.}~\bibnamefont {Dequal}}, \bibinfo
  {author} {\bibfnamefont {P.}~\bibnamefont {Villoresi}}, \ and\ \bibinfo
  {author} {\bibfnamefont {G.}~\bibnamefont {Vallone}},\ }\href@noop {}
  {\bibfield  {journal} {\bibinfo  {journal} {Phys. Rev. Lett.}\ }\textbf
  {\bibinfo {volume} {121}},\ \bibinfo {pages} {230501} (\bibinfo {year}
  {2018})}\BibitemShut {NoStop}%
\bibitem [{\citenamefont {Bolduc}\ \emph {et~al.}(2016)\citenamefont {Bolduc},
  \citenamefont {Gariepy},\ and\ \citenamefont {Leach}}]{bolduc2016direct}%
  \BibitemOpen
  \bibfield  {author} {\bibinfo {author} {\bibfnamefont {E.}~\bibnamefont
  {Bolduc}}, \bibinfo {author} {\bibfnamefont {G.}~\bibnamefont {Gariepy}}, \
  and\ \bibinfo {author} {\bibfnamefont {J.}~\bibnamefont {Leach}},\
  }\href@noop {} {\bibfield  {journal} {\bibinfo  {journal} {Nature
  communications}\ }\textbf {\bibinfo {volume} {7}},\ \bibinfo {pages} {10439}
  (\bibinfo {year} {2016})}\BibitemShut {NoStop}%
\bibitem [{\citenamefont {Pan}\ \emph {et~al.}(2019)\citenamefont {Pan},
  \citenamefont {Xu}, \citenamefont {Kedem}, \citenamefont {Wang},
  \citenamefont {Chen}, \citenamefont {Jan}, \citenamefont {Sun}, \citenamefont
  {Xu}, \citenamefont {Han}, \citenamefont {Li} \emph
  {et~al.}}]{pan2019direct}%
  \BibitemOpen
  \bibfield  {author} {\bibinfo {author} {\bibfnamefont {W.-W.}\ \bibnamefont
  {Pan}}, \bibinfo {author} {\bibfnamefont {X.-Y.}\ \bibnamefont {Xu}},
  \bibinfo {author} {\bibfnamefont {Y.}~\bibnamefont {Kedem}}, \bibinfo
  {author} {\bibfnamefont {Q.-Q.}\ \bibnamefont {Wang}}, \bibinfo {author}
  {\bibfnamefont {Z.}~\bibnamefont {Chen}}, \bibinfo {author} {\bibfnamefont
  {M.}~\bibnamefont {Jan}}, \bibinfo {author} {\bibfnamefont {K.}~\bibnamefont
  {Sun}}, \bibinfo {author} {\bibfnamefont {J.-S.}\ \bibnamefont {Xu}},
  \bibinfo {author} {\bibfnamefont {Y.-J.}\ \bibnamefont {Han}}, \bibinfo
  {author} {\bibfnamefont {C.-F.}\ \bibnamefont {Li}},  \emph {et~al.},\
  }\href@noop {} {\bibfield  {journal} {\bibinfo  {journal} {Physical review
  letters}\ }\textbf {\bibinfo {volume} {123}},\ \bibinfo {pages} {150402}
  (\bibinfo {year} {2019})}\BibitemShut {NoStop}%
\bibitem [{\citenamefont {Bennett}\ \emph {et~al.}(1993)\citenamefont
  {Bennett}, \citenamefont {Brassard}, \citenamefont {Cr{\'e}peau},
  \citenamefont {Jozsa}, \citenamefont {Peres},\ and\ \citenamefont
  {Wootters}}]{bennett1993teleporting}%
  \BibitemOpen
  \bibfield  {author} {\bibinfo {author} {\bibfnamefont {C.~H.}\ \bibnamefont
  {Bennett}}, \bibinfo {author} {\bibfnamefont {G.}~\bibnamefont {Brassard}},
  \bibinfo {author} {\bibfnamefont {C.}~\bibnamefont {Cr{\'e}peau}}, \bibinfo
  {author} {\bibfnamefont {R.}~\bibnamefont {Jozsa}}, \bibinfo {author}
  {\bibfnamefont {A.}~\bibnamefont {Peres}}, \ and\ \bibinfo {author}
  {\bibfnamefont {W.~K.}\ \bibnamefont {Wootters}},\ }\href@noop {} {\bibfield
  {journal} {\bibinfo  {journal} {Physical review letters}\ }\textbf {\bibinfo
  {volume} {70}},\ \bibinfo {pages} {1895} (\bibinfo {year}
  {1993})}\BibitemShut {NoStop}%
\bibitem [{\citenamefont {Kimble}(2008)}]{kimble2008quantum}%
  \BibitemOpen
  \bibfield  {author} {\bibinfo {author} {\bibfnamefont {H.~J.}\ \bibnamefont
  {Kimble}},\ }\href@noop {} {\bibfield  {journal} {\bibinfo  {journal}
  {Nature}\ }\textbf {\bibinfo {volume} {453}},\ \bibinfo {pages} {1023}
  (\bibinfo {year} {2008})}\BibitemShut {NoStop}%
\bibitem [{\citenamefont {Gottesman}\ and\ \citenamefont
  {Chuang}(1999)}]{gottesman1999demonstrating}%
  \BibitemOpen
  \bibfield  {author} {\bibinfo {author} {\bibfnamefont {D.}~\bibnamefont
  {Gottesman}}\ and\ \bibinfo {author} {\bibfnamefont {I.~L.}\ \bibnamefont
  {Chuang}},\ }\href@noop {} {\bibfield  {journal} {\bibinfo  {journal}
  {Nature}\ }\textbf {\bibinfo {volume} {402}},\ \bibinfo {pages} {390}
  (\bibinfo {year} {1999})}\BibitemShut {NoStop}%
\bibitem [{\citenamefont {Raussendorf}\ and\ \citenamefont
  {Briegel}(2001)}]{raussendorf2001one}%
  \BibitemOpen
  \bibfield  {author} {\bibinfo {author} {\bibfnamefont {R.}~\bibnamefont
  {Raussendorf}}\ and\ \bibinfo {author} {\bibfnamefont {H.~J.}\ \bibnamefont
  {Briegel}},\ }\href@noop {} {\bibfield  {journal} {\bibinfo  {journal}
  {Physical Review Letters}\ }\textbf {\bibinfo {volume} {86}},\ \bibinfo
  {pages} {5188} (\bibinfo {year} {2001})}\BibitemShut {NoStop}%
\bibitem [{\citenamefont {Okamoto}\ \emph {et~al.}(2009)\citenamefont
  {Okamoto}, \citenamefont {O'brien}, \citenamefont {Hofmann}, \citenamefont
  {Nagata}, \citenamefont {Sasaki},\ and\ \citenamefont
  {Takeuchi}}]{okamoto2009entanglement}%
  \BibitemOpen
  \bibfield  {author} {\bibinfo {author} {\bibfnamefont {R.}~\bibnamefont
  {Okamoto}}, \bibinfo {author} {\bibfnamefont {J.~L.}\ \bibnamefont
  {O'brien}}, \bibinfo {author} {\bibfnamefont {H.~F.}\ \bibnamefont
  {Hofmann}}, \bibinfo {author} {\bibfnamefont {T.}~\bibnamefont {Nagata}},
  \bibinfo {author} {\bibfnamefont {K.}~\bibnamefont {Sasaki}}, \ and\ \bibinfo
  {author} {\bibfnamefont {S.}~\bibnamefont {Takeuchi}},\ }\href@noop {}
  {\bibfield  {journal} {\bibinfo  {journal} {Science}\ }\textbf {\bibinfo
  {volume} {323}},\ \bibinfo {pages} {483} (\bibinfo {year}
  {2009})}\BibitemShut {NoStop}%
\bibitem [{\citenamefont {Greenberger}\ \emph {et~al.}(1990)\citenamefont
  {Greenberger}, \citenamefont {Horne}, \citenamefont {Shimony},\ and\
  \citenamefont {Zeilinger}}]{greenberger1990bell}%
  \BibitemOpen
  \bibfield  {author} {\bibinfo {author} {\bibfnamefont {D.~M.}\ \bibnamefont
  {Greenberger}}, \bibinfo {author} {\bibfnamefont {M.~A.}\ \bibnamefont
  {Horne}}, \bibinfo {author} {\bibfnamefont {A.}~\bibnamefont {Shimony}}, \
  and\ \bibinfo {author} {\bibfnamefont {A.}~\bibnamefont {Zeilinger}},\
  }\href@noop {} {\bibfield  {journal} {\bibinfo  {journal} {American Journal
  of Physics}\ }\textbf {\bibinfo {volume} {58}},\ \bibinfo {pages} {1131}
  (\bibinfo {year} {1990})}\BibitemShut {NoStop}%
\bibitem [{\citenamefont {Bouwmeester}\ \emph {et~al.}(1997)\citenamefont
  {Bouwmeester}, \citenamefont {Pan}, \citenamefont {Mattle}, \citenamefont
  {Eibl}, \citenamefont {Weinfurter},\ and\ \citenamefont
  {Zeilinger}}]{bouwmeester1997experimental}%
  \BibitemOpen
  \bibfield  {author} {\bibinfo {author} {\bibfnamefont {D.}~\bibnamefont
  {Bouwmeester}}, \bibinfo {author} {\bibfnamefont {J.-W.}\ \bibnamefont
  {Pan}}, \bibinfo {author} {\bibfnamefont {K.}~\bibnamefont {Mattle}},
  \bibinfo {author} {\bibfnamefont {M.}~\bibnamefont {Eibl}}, \bibinfo {author}
  {\bibfnamefont {H.}~\bibnamefont {Weinfurter}}, \ and\ \bibinfo {author}
  {\bibfnamefont {A.}~\bibnamefont {Zeilinger}},\ }\href@noop {} {\bibfield
  {journal} {\bibinfo  {journal} {Nature}\ }\textbf {\bibinfo {volume} {390}},\
  \bibinfo {pages} {575} (\bibinfo {year} {1997})}\BibitemShut {NoStop}%
\bibitem [{\citenamefont {Zhao}\ \emph {et~al.}(2004)\citenamefont {Zhao},
  \citenamefont {Chen}, \citenamefont {Zhang}, \citenamefont {Yang},
  \citenamefont {Briegel},\ and\ \citenamefont {Pan}}]{zhao2004experimental}%
  \BibitemOpen
  \bibfield  {author} {\bibinfo {author} {\bibfnamefont {Z.}~\bibnamefont
  {Zhao}}, \bibinfo {author} {\bibfnamefont {Y.-A.}\ \bibnamefont {Chen}},
  \bibinfo {author} {\bibfnamefont {A.-N.}\ \bibnamefont {Zhang}}, \bibinfo
  {author} {\bibfnamefont {T.}~\bibnamefont {Yang}}, \bibinfo {author}
  {\bibfnamefont {H.~J.}\ \bibnamefont {Briegel}}, \ and\ \bibinfo {author}
  {\bibfnamefont {J.-W.}\ \bibnamefont {Pan}},\ }\href@noop {} {\bibfield
  {journal} {\bibinfo  {journal} {Nature}\ }\textbf {\bibinfo {volume} {430}},\
  \bibinfo {pages} {54} (\bibinfo {year} {2004})}\BibitemShut {NoStop}%
\bibitem [{\citenamefont {Zhang}\ \emph {et~al.}(2006)\citenamefont {Zhang},
  \citenamefont {Goebel}, \citenamefont {Wagenknecht}, \citenamefont {Chen},
  \citenamefont {Zhao}, \citenamefont {Yang}, \citenamefont {Mair},
  \citenamefont {Schmiedmayer},\ and\ \citenamefont
  {Pan}}]{zhang2006experimental}%
  \BibitemOpen
  \bibfield  {author} {\bibinfo {author} {\bibfnamefont {Q.}~\bibnamefont
  {Zhang}}, \bibinfo {author} {\bibfnamefont {A.}~\bibnamefont {Goebel}},
  \bibinfo {author} {\bibfnamefont {C.}~\bibnamefont {Wagenknecht}}, \bibinfo
  {author} {\bibfnamefont {Y.-A.}\ \bibnamefont {Chen}}, \bibinfo {author}
  {\bibfnamefont {B.}~\bibnamefont {Zhao}}, \bibinfo {author} {\bibfnamefont
  {T.}~\bibnamefont {Yang}}, \bibinfo {author} {\bibfnamefont {A.}~\bibnamefont
  {Mair}}, \bibinfo {author} {\bibfnamefont {J.}~\bibnamefont {Schmiedmayer}},
  \ and\ \bibinfo {author} {\bibfnamefont {J.-W.}\ \bibnamefont {Pan}},\
  }\href@noop {} {\bibfield  {journal} {\bibinfo  {journal} {Nature Physics}\
  }\textbf {\bibinfo {volume} {2}},\ \bibinfo {pages} {678} (\bibinfo {year}
  {2006})}\BibitemShut {NoStop}%
\bibitem [{\citenamefont {Wang}\ \emph {et~al.}(2015)\citenamefont {Wang},
  \citenamefont {Cai}, \citenamefont {Su}, \citenamefont {Chen}, \citenamefont
  {Wu}, \citenamefont {Li}, \citenamefont {Liu}, \citenamefont {Lu},\ and\
  \citenamefont {Pan}}]{wang2015quantum}%
  \BibitemOpen
  \bibfield  {author} {\bibinfo {author} {\bibfnamefont {X.-L.}\ \bibnamefont
  {Wang}}, \bibinfo {author} {\bibfnamefont {X.-D.}\ \bibnamefont {Cai}},
  \bibinfo {author} {\bibfnamefont {Z.-E.}\ \bibnamefont {Su}}, \bibinfo
  {author} {\bibfnamefont {M.-C.}\ \bibnamefont {Chen}}, \bibinfo {author}
  {\bibfnamefont {D.}~\bibnamefont {Wu}}, \bibinfo {author} {\bibfnamefont
  {L.}~\bibnamefont {Li}}, \bibinfo {author} {\bibfnamefont {N.-L.}\
  \bibnamefont {Liu}}, \bibinfo {author} {\bibfnamefont {C.-Y.}\ \bibnamefont
  {Lu}}, \ and\ \bibinfo {author} {\bibfnamefont {J.-W.}\ \bibnamefont {Pan}},\
  }\href@noop {} {\bibfield  {journal} {\bibinfo  {journal} {Nature}\ }\textbf
  {\bibinfo {volume} {518}},\ \bibinfo {pages} {516} (\bibinfo {year}
  {2015})}\BibitemShut {NoStop}%
\bibitem [{\citenamefont {Krauter}\ \emph {et~al.}(2013)\citenamefont
  {Krauter}, \citenamefont {Salart}, \citenamefont {Muschik}, \citenamefont
  {Petersen}, \citenamefont {Shen}, \citenamefont {Fernholz},\ and\
  \citenamefont {Polzik}}]{krauter2013deterministic}%
  \BibitemOpen
  \bibfield  {author} {\bibinfo {author} {\bibfnamefont {H.}~\bibnamefont
  {Krauter}}, \bibinfo {author} {\bibfnamefont {D.}~\bibnamefont {Salart}},
  \bibinfo {author} {\bibfnamefont {C.}~\bibnamefont {Muschik}}, \bibinfo
  {author} {\bibfnamefont {J.~M.}\ \bibnamefont {Petersen}}, \bibinfo {author}
  {\bibfnamefont {H.}~\bibnamefont {Shen}}, \bibinfo {author} {\bibfnamefont
  {T.}~\bibnamefont {Fernholz}}, \ and\ \bibinfo {author} {\bibfnamefont
  {E.~S.}\ \bibnamefont {Polzik}},\ }\href@noop {} {\bibfield  {journal}
  {\bibinfo  {journal} {Nature Physics}\ }\textbf {\bibinfo {volume} {9}},\
  \bibinfo {pages} {400} (\bibinfo {year} {2013})}\BibitemShut {NoStop}%
\bibitem [{\citenamefont {Riebe}\ \emph {et~al.}(2004)\citenamefont {Riebe},
  \citenamefont {H{\"a}ffner}, \citenamefont {Roos}, \citenamefont
  {H{\"a}nsel}, \citenamefont {Benhelm}, \citenamefont {Lancaster},
  \citenamefont {K{\"o}rber}, \citenamefont {Becher}, \citenamefont
  {Schmidt-Kaler}, \citenamefont {James} \emph
  {et~al.}}]{riebe2004deterministic}%
  \BibitemOpen
  \bibfield  {author} {\bibinfo {author} {\bibfnamefont {M.}~\bibnamefont
  {Riebe}}, \bibinfo {author} {\bibfnamefont {H.}~\bibnamefont {H{\"a}ffner}},
  \bibinfo {author} {\bibfnamefont {C.}~\bibnamefont {Roos}}, \bibinfo {author}
  {\bibfnamefont {W.}~\bibnamefont {H{\"a}nsel}}, \bibinfo {author}
  {\bibfnamefont {J.}~\bibnamefont {Benhelm}}, \bibinfo {author} {\bibfnamefont
  {G.}~\bibnamefont {Lancaster}}, \bibinfo {author} {\bibfnamefont
  {T.}~\bibnamefont {K{\"o}rber}}, \bibinfo {author} {\bibfnamefont
  {C.}~\bibnamefont {Becher}}, \bibinfo {author} {\bibfnamefont
  {F.}~\bibnamefont {Schmidt-Kaler}}, \bibinfo {author} {\bibfnamefont
  {D.}~\bibnamefont {James}},  \emph {et~al.},\ }\href@noop {} {\bibfield
  {journal} {\bibinfo  {journal} {Nature}\ }\textbf {\bibinfo {volume} {429}},\
  \bibinfo {pages} {734} (\bibinfo {year} {2004})}\BibitemShut {NoStop}%
\bibitem [{\citenamefont {Barrett}\ \emph {et~al.}(2004)\citenamefont
  {Barrett}, \citenamefont {Chiaverini}, \citenamefont {Schaetz}, \citenamefont
  {Britton}, \citenamefont {Itano}, \citenamefont {Jost}, \citenamefont
  {Knill}, \citenamefont {Langer}, \citenamefont {Leibfried}, \citenamefont
  {Ozeri} \emph {et~al.}}]{barrett2004deterministic}%
  \BibitemOpen
  \bibfield  {author} {\bibinfo {author} {\bibfnamefont {M.}~\bibnamefont
  {Barrett}}, \bibinfo {author} {\bibfnamefont {J.}~\bibnamefont {Chiaverini}},
  \bibinfo {author} {\bibfnamefont {T.}~\bibnamefont {Schaetz}}, \bibinfo
  {author} {\bibfnamefont {J.}~\bibnamefont {Britton}}, \bibinfo {author}
  {\bibfnamefont {W.}~\bibnamefont {Itano}}, \bibinfo {author} {\bibfnamefont
  {J.}~\bibnamefont {Jost}}, \bibinfo {author} {\bibfnamefont {E.}~\bibnamefont
  {Knill}}, \bibinfo {author} {\bibfnamefont {C.}~\bibnamefont {Langer}},
  \bibinfo {author} {\bibfnamefont {D.}~\bibnamefont {Leibfried}}, \bibinfo
  {author} {\bibfnamefont {R.}~\bibnamefont {Ozeri}},  \emph {et~al.},\
  }\href@noop {} {\bibfield  {journal} {\bibinfo  {journal} {Nature}\ }\textbf
  {\bibinfo {volume} {429}},\ \bibinfo {pages} {737} (\bibinfo {year}
  {2004})}\BibitemShut {NoStop}%
\bibitem [{\citenamefont {Pfaff}\ \emph {et~al.}(2014)\citenamefont {Pfaff},
  \citenamefont {Hensen}, \citenamefont {Bernien}, \citenamefont {van Dam},
  \citenamefont {Blok}, \citenamefont {Taminiau}, \citenamefont {Tiggelman},
  \citenamefont {Schouten}, \citenamefont {Markham}, \citenamefont {Twitchen}
  \emph {et~al.}}]{pfaff2014unconditional}%
  \BibitemOpen
  \bibfield  {author} {\bibinfo {author} {\bibfnamefont {W.}~\bibnamefont
  {Pfaff}}, \bibinfo {author} {\bibfnamefont {B.}~\bibnamefont {Hensen}},
  \bibinfo {author} {\bibfnamefont {H.}~\bibnamefont {Bernien}}, \bibinfo
  {author} {\bibfnamefont {S.~B.}\ \bibnamefont {van Dam}}, \bibinfo {author}
  {\bibfnamefont {M.~S.}\ \bibnamefont {Blok}}, \bibinfo {author}
  {\bibfnamefont {T.~H.}\ \bibnamefont {Taminiau}}, \bibinfo {author}
  {\bibfnamefont {M.~J.}\ \bibnamefont {Tiggelman}}, \bibinfo {author}
  {\bibfnamefont {R.~N.}\ \bibnamefont {Schouten}}, \bibinfo {author}
  {\bibfnamefont {M.}~\bibnamefont {Markham}}, \bibinfo {author} {\bibfnamefont
  {D.~J.}\ \bibnamefont {Twitchen}},  \emph {et~al.},\ }\href@noop {}
  {\bibfield  {journal} {\bibinfo  {journal} {Science}\ }\textbf {\bibinfo
  {volume} {345}},\ \bibinfo {pages} {532} (\bibinfo {year}
  {2014})}\BibitemShut {NoStop}%
\bibitem [{\citenamefont {Steffen}\ \emph {et~al.}(2013)\citenamefont
  {Steffen}, \citenamefont {Salathe}, \citenamefont {Oppliger}, \citenamefont
  {Kurpiers}, \citenamefont {Baur}, \citenamefont {Lang}, \citenamefont
  {Eichler}, \citenamefont {Puebla-Hellmann}, \citenamefont {Fedorov},\ and\
  \citenamefont {Wallraff}}]{steffen2013deterministic}%
  \BibitemOpen
  \bibfield  {author} {\bibinfo {author} {\bibfnamefont {L.}~\bibnamefont
  {Steffen}}, \bibinfo {author} {\bibfnamefont {Y.}~\bibnamefont {Salathe}},
  \bibinfo {author} {\bibfnamefont {M.}~\bibnamefont {Oppliger}}, \bibinfo
  {author} {\bibfnamefont {P.}~\bibnamefont {Kurpiers}}, \bibinfo {author}
  {\bibfnamefont {M.}~\bibnamefont {Baur}}, \bibinfo {author} {\bibfnamefont
  {C.}~\bibnamefont {Lang}}, \bibinfo {author} {\bibfnamefont {C.}~\bibnamefont
  {Eichler}}, \bibinfo {author} {\bibfnamefont {G.}~\bibnamefont
  {Puebla-Hellmann}}, \bibinfo {author} {\bibfnamefont {A.}~\bibnamefont
  {Fedorov}}, \ and\ \bibinfo {author} {\bibfnamefont {A.}~\bibnamefont
  {Wallraff}},\ }\href@noop {} {\bibfield  {journal} {\bibinfo  {journal}
  {Nature}\ }\textbf {\bibinfo {volume} {500}},\ \bibinfo {pages} {319}
  (\bibinfo {year} {2013})}\BibitemShut {NoStop}%
\bibitem [{\citenamefont {Zhang}\ \emph {et~al.}(2015)\citenamefont {Zhang},
  \citenamefont {Huang}, \citenamefont {Wang}, \citenamefont {Liu},
  \citenamefont {Li},\ and\ \citenamefont {Guo}}]{zhang2015experimental}%
  \BibitemOpen
  \bibfield  {author} {\bibinfo {author} {\bibfnamefont {C.}~\bibnamefont
  {Zhang}}, \bibinfo {author} {\bibfnamefont {Y.-F.}\ \bibnamefont {Huang}},
  \bibinfo {author} {\bibfnamefont {Z.}~\bibnamefont {Wang}}, \bibinfo {author}
  {\bibfnamefont {B.-H.}\ \bibnamefont {Liu}}, \bibinfo {author} {\bibfnamefont
  {C.-F.}\ \bibnamefont {Li}}, \ and\ \bibinfo {author} {\bibfnamefont {G.-C.}\
  \bibnamefont {Guo}},\ }\href@noop {} {\bibfield  {journal} {\bibinfo
  {journal} {Physical review letters}\ }\textbf {\bibinfo {volume} {115}},\
  \bibinfo {pages} {260402} (\bibinfo {year} {2015})}\BibitemShut {NoStop}%
\bibitem [{\citenamefont {Cramer}\ \emph {et~al.}(2010)\citenamefont {Cramer},
  \citenamefont {Plenio}, \citenamefont {Flammia}, \citenamefont {Somma},
  \citenamefont {Gross}, \citenamefont {Bartlett}, \citenamefont
  {Landon-Cardinal}, \citenamefont {Poulin},\ and\ \citenamefont
  {Liu}}]{cramer2010efficient}%
  \BibitemOpen
  \bibfield  {author} {\bibinfo {author} {\bibfnamefont {M.}~\bibnamefont
  {Cramer}}, \bibinfo {author} {\bibfnamefont {M.~B.}\ \bibnamefont {Plenio}},
  \bibinfo {author} {\bibfnamefont {S.~T.}\ \bibnamefont {Flammia}}, \bibinfo
  {author} {\bibfnamefont {R.}~\bibnamefont {Somma}}, \bibinfo {author}
  {\bibfnamefont {D.}~\bibnamefont {Gross}}, \bibinfo {author} {\bibfnamefont
  {S.~D.}\ \bibnamefont {Bartlett}}, \bibinfo {author} {\bibfnamefont
  {O.}~\bibnamefont {Landon-Cardinal}}, \bibinfo {author} {\bibfnamefont
  {D.}~\bibnamefont {Poulin}}, \ and\ \bibinfo {author} {\bibfnamefont {Y.-K.}\
  \bibnamefont {Liu}},\ }\href@noop {} {\bibfield  {journal} {\bibinfo
  {journal} {Nature communications}\ }\textbf {\bibinfo {volume} {1}},\
  \bibinfo {pages} {1} (\bibinfo {year} {2010})}\BibitemShut {NoStop}%
\bibitem [{\citenamefont {Torlai}\ \emph {et~al.}(2018)\citenamefont {Torlai},
  \citenamefont {Mazzola}, \citenamefont {Carrasquilla}, \citenamefont
  {Troyer}, \citenamefont {Melko},\ and\ \citenamefont
  {Carleo}}]{torlai2018neural}%
  \BibitemOpen
  \bibfield  {author} {\bibinfo {author} {\bibfnamefont {G.}~\bibnamefont
  {Torlai}}, \bibinfo {author} {\bibfnamefont {G.}~\bibnamefont {Mazzola}},
  \bibinfo {author} {\bibfnamefont {J.}~\bibnamefont {Carrasquilla}}, \bibinfo
  {author} {\bibfnamefont {M.}~\bibnamefont {Troyer}}, \bibinfo {author}
  {\bibfnamefont {R.}~\bibnamefont {Melko}}, \ and\ \bibinfo {author}
  {\bibfnamefont {G.}~\bibnamefont {Carleo}},\ }\href@noop {} {\bibfield
  {journal} {\bibinfo  {journal} {Nature Physics}\ }\textbf {\bibinfo {volume}
  {14}},\ \bibinfo {pages} {447} (\bibinfo {year} {2018})}\BibitemShut
  {NoStop}%
\bibitem [{\citenamefont {Gross}\ \emph {et~al.}(2010)\citenamefont {Gross},
  \citenamefont {Liu}, \citenamefont {Flammia}, \citenamefont {Becker},\ and\
  \citenamefont {Eisert}}]{gross2010quantum}%
  \BibitemOpen
  \bibfield  {author} {\bibinfo {author} {\bibfnamefont {D.}~\bibnamefont
  {Gross}}, \bibinfo {author} {\bibfnamefont {Y.-K.}\ \bibnamefont {Liu}},
  \bibinfo {author} {\bibfnamefont {S.~T.}\ \bibnamefont {Flammia}}, \bibinfo
  {author} {\bibfnamefont {S.}~\bibnamefont {Becker}}, \ and\ \bibinfo {author}
  {\bibfnamefont {J.}~\bibnamefont {Eisert}},\ }\href@noop {} {\bibfield
  {journal} {\bibinfo  {journal} {Physical review letters}\ }\textbf {\bibinfo
  {volume} {105}},\ \bibinfo {pages} {150401} (\bibinfo {year}
  {2010})}\BibitemShut {NoStop}%
\end{thebibliography}%

\end{document}